\documentclass{article}

\newcounter{nb_tests}
\newcounter{nb_explainers}
\newcounter{nb_papers}
\PassOptionsToPackage{numbers, compress}{natbib} 

    \usepackage[preprint]{neurips_data_2022}

\usepackage[utf8]{inputenc} 
\usepackage[T1]{fontenc}    
\usepackage{hyperref}       
\usepackage{url}            
\usepackage{booktabs}       
\usepackage{amsfonts}       
\usepackage{nicefrac}       
\usepackage{microtype}      
\usepackage{xcolor}         
\usepackage{graphicx}       

\title{Compare-xAI: Toward Unifying Functional Testing Methods for Post-hoc XAI Algorithms into an Interactive and Multi-dimensional Benchmark}

\author{
 Mohamed Karim Belaid 
 \\
 IDIADA Fahrzeugtechnik GmbH \\
  Munich, Germany\\
 \texttt{Karim.Belaid@idiada.com} \\
  \And
  Eyke H{\"u}llermeier \\
  University of Munich (LMU) \\
  Munich, Germany \\ 
  \texttt{eyke@if.lmu.de} \\
  \AND
  Maximilian Rabus \\
  Dr.\ Ing.\ h.c.\ F.\ Porsche AG \\
  Stuttgart, Germany \\
  \texttt{maximilian.rabus2@porsche.de} \\
  \And
  Ralf Krestel \\
  ZBW - Leibniz Centre for Economics \\
  \& Kiel University \\
  Kiel, Germany \\
  \texttt{r.krestel@zbw.eu} \\
}

\graphicspath{{img/}} 

\begin{document}

\maketitle

\begin{abstract}
In recent years, Explainable AI (xAI) attracted a lot of attention as various countries turned explanations into a legal right. xAI allows for improving models beyond the accuracy metric by, e.g., debugging the learned pattern and demystifying the AI's behavior. The widespread use of xAI brought new challenges. On the one hand, the number of published xAI algorithms underwent a boom, and it became difficult for practitioners to select the right tool. On the other hand, some experiments did highlight how easy data scientists could misuse xAI algorithms and misinterpret their results. To tackle the issue of comparing and correctly using feature importance xAI algorithms, 
we propose Compare-xAI, a benchmark that unifies all exclusive functional testing methods applied to xAI algorithms. We propose a selection protocol to shortlist non-redundant functional tests from the literature, i.e., each targeting a specific end-user requirement in explaining a model. The benchmark encapsulates the complexity of evaluating xAI methods into a hierarchical scoring of three levels, namely, targeting three end-user groups: researchers, practitioners, and laymen in xAI. The most detailed level provides one score per test. The second level regroups tests into five categories (fidelity, fragility, stability, simplicity, and stress tests). The last level is the aggregated comprehensibility score, which encapsulates the ease of correctly interpreting the algorithm's output in one easy to compare value. 
Compare-xAI's interactive user interface helps mitigate errors in interpreting xAI results by quickly listing the recommended xAI solutions for each ML task and their current limitations. 
The benchmark is made available at \url{https://karim-53.github.io/cxai/}
\end{abstract}

\section{Introduction}
xAI algorithms are a set of approaches toward understanding black-box models. In recent years, xAI algorithms helped debug manifold issues in ML models, such as exposing underlying wrong patterns in classifying objects~\citep{ribeiro2016should} or highlighting inequality and bias in decisions~\citep{dressel2018accuracy}. Moreover, given its essential impact on society, legislation in several countries now includes the ``Right to explanation''~\citep{goodman2017european} fulfilled by the various xAI tools available in the literature. 
It is indeed difficult to define the best xAI solution given the number of known evaluation metrics. 
Moreover, the long evolutionary history of specific xAI methods makes it even more difficult to evaluate each version. The Shapley values are an excellent example of this challenge. Sundararajan et al. did state that ``\ldots the functional forms of the Shapley value\ldots are sufficiently complex as to prevent direct understanding\ldots''~\citep{sundararajan2020many}.
Indeed, going through the theoretical background of Shapley values~\citep{shapley1953quota}, its multiple approximations~\citep{vstrumbelj2014explaining,lundberg2017unified}, 
generalizations~\citep{sundararajan2017axiomatic,sundararajan2020many} and final implementations~\citep{,lundberg2018consistent,staniak2018explanations} adapted to the AI field might mislead the end-user on the capability of the available tools.

\paragraph{Resulting challenges.} Consequently, data scientists face considerable difficulties in accurately evaluating each xAI algorithm and remaining up-to-date on its evolution. This issue yields a clearly visible symptom known as the illusion of explanatory depth~\citep{rozenblit2002misunderstood} in interpreting xAI results~\citep{chromik2021think} as it has been confirmed that data scientists are prone to misuse interpretability tools~\citep{kaur2020interpreting}. Many researchers did address this question by stressing the importance of structuring and documenting xAI algorithms~\citep{leavitt2020towards,kearns2019ethical}, i.e., by highlighting the target end-users of the algorithm, its capability, limitations, and vulnerabilities. Finally, they recommend using quantitative metrics to make claims about explainability.
\paragraph{Functional testing as a solution to stated recommendations.}
Functional testing aim to verify the end-user's requirement on the xAI algorithm by performing end-to-end tests in a black-box fashion. In other words, every functional test apply the xAI algorithm on a frozen AI model to verify if the output corresponds to the explanation expected by data scientists. A functional test could verify that the explanation accurately reflect the AI model (Fidelity), that it is not sensitive to adversarial attacks (Fragility), that it is stable to small variation in the model (Stability), etc. Functional testing remains unfortunately sparsely used in literature and, thus, provides only a partial evaluation.

Given the unsolved burden of evaluating and correctly interpreting xAI results, we propose Compare-xAI that mitigates these two issues (benchmark xAI results and the illusion of explanatory depth during the interpretation of results) by addressing three research questions:

\begin{enumerate}
    \item How to select exclusive functional tests from those proposed in the literature?
    \item How to score xAI algorithms in a simple way despite the multitude of evaluation dimensions? 
    \item How to reduce data scientists' potential misuse of the xAI algorithms?
\end{enumerate}
The stated questions are resolved as follows: 
In Section~\ref{sec:main}, we propose a benchmark implementation easily scalable to new xAI algorithms and new functional tests.
In Section~\ref{sub:test_protocol}, we propose a selection protocol for the quantitative evaluation of xAI algorithms. It is then applied to shortlist a selection of exclusive tests, each targeting a distinct end-user requirement.
In Section~\ref{sub:experiment_protocol}, we explain the experiments' protocol to mimic the end-user's behavior.
In Section~\ref{sub:scoring_protocol}, we propose an intuitive scoring method that scales in detail with the level of expertise of the data scientist: Layman data scientists are invited to manipulate one global score named comprehensibility. Practitioners are invited to compare xAI algorithms given five scores representing five subcategories of the comprehensibility metric. Finally, researchers are invited to study the detailed report (one score per test).
In Section~\ref{sec:interface}, we propose a user interface that encapsulates the benchmark's results. We seek to minimize the potential misuse of xAI algorithms by offering quick access to the limitation of each xAI algorithm.
Finally, Section~\ref{sec:limitation} is dedicated to the theoretical and practical limitations of the benchmark.

\section{Related Work}\label{sec:related}
This section is a survey for xAI evaluation methods. It contains examples contrasting the difference between functional tests and portability tests. Following that, we examine some attempts to regroup them into surveys or benchmarks.

\subsection{xAI Evaluation Methods: Functional Tests vs.\ Portability Tests}
\label{sub:tests}

Researcher in the xAI field often propose a new method along with a set of functional or portability tests that outline the contrast between former work and their contribution.
\paragraph{Functional tests.} Functional testing is a popular testing technique for software engineers.
The following definition is adapted from the software engineering field to our intended usage in machine learning~\citep{beizer1995black}. 
Functional tests are created by testers with no specific knowledge of the algorithm's internal modules, i.e., not the developers themselves. Therefore, the algorithm is considered a black-box and is executed from end to end. 
Each functional test is intended to verify an end-user requirement on the xAI algorithm rather than a specific internal module. Thus, functional tests share the advantage of being able to test different algorithms. 
On the other hand, failed tests do not inform about the location of the errors but rather attribute it to the entire algorithm. Functional test for xAI algorithms usually exploit tabular synthetic data and few input features, e.g., considering the ``cough and fever'' test~\citep{lundberg2018consistent}, The xAI algorithm is expected to detect symmetry between the two binary features. Simple examples showcase the undeniable limit of certain xAI methods.
Nevertheless, specific tests could use real-world data. A good example is the MNIST dataset~\citep{lecun2010mnist} used as a counterexample for the dummy axiom~\citep{covert2020understanding}: Since edge pixels are always black, a multi-layer perceptron will learn not to rely on these constant pixels. As a consequence, the xAI algorithm should confirm that the AI does not use these pixels.
Papers proposing new xAI methods remain too short to list all known tests. Furthermore, some of the highlighted issues might be fixed without any publication.
\paragraph{Portability Tests.} Portability tests for xAI algorithms evaluate real-word models and demonstrate the robustness of the xAI algorithm against multiple challenges at once (noise, correlated inputs, large inputs, etc.). They are used to claim the potential broad usage of one xAI method rather than demonstrating the quality of the explanation. An example is the verification of the portability across recommendation tasks~\citep{tsang2020does}.

\paragraph{Navigating the ocean of tests remains itself a huge challenge.} First, many examples in the literature are portability tests which makes comparison between xAI algorithms complex. 
Second, tests could be redundant to emphasize the frequent occurrence of an issue, e.g., testing interaction detection with different transparent models~\citep{tsang2020does}.
Third, researchers could argue the correctness of specific functional tests' ground truth, e.g., causal explanation of the Shapley values~\citep{kumar2020problems} has been considered false in certain research~\citep{sundararajan2020many}.

Given the tremendous amount of xAI algorithms and dedicated metrics, surveys~\citep{mohseni2018multidisciplinary,tsang2021interpretable,angelov2021explainable,rozemberczki2022shapley} have trouble providing an in-depth analysis of each algorithm and cannot cope with ongoing implementation updates. Nevertheless, Molnar's online book distinguishes itself with a continuously updated survey about xAI~\citep{molnar2020interpretable}. The initiative of a real-time survey faced great success and acceptance from the data science community. 

\subsection{Benchmark for xAI Algorithms}
There are specialized benchmarks in the literature, like the SVEA benchmark~\citep{sattarzadeh2021svea}. The latter focuses on computer vision tasks and proposes faster evaluations based on the small mnist-1D dataset~\citep{greydanus2020scaling}. Another benchmark utilizes exclusively human evaluation to assess xAI algorithms on real-world tasks~\citep{mohseni2020benchmark}. On the one hand, benchmarking using computer vision and NLP models permits to measure the real success of an xAI tool in helping end-users even though human evaluation could be considered subjective and more costly to obtain. On the other hand, evaluation using real-world tasks does not allow debugging the xAI algorithm, i.e., two algorithms might fail to explain one black-box model for two different reasons.

xAI-Bench~\citep{liu2021synthetic} 
evaluates each xAI algorithm on five metrics. 
Faithfulness measures the Pearson correlation between the feature importance and the approximate marginal contribution of each feature. Of course, one could argue that the ground truth explanation of a model could be slightly different from the marginal contribution of each feature on the observed dataset. 
The same argument holds for the monotonicity, infidelity, and GT-Shapley metrics. They define a ground truth output, that is, a ``better'' xAI algorithm is the one outputting a result that is closer to the ground truth. In contrast, functional tests discussed in previous paragraphs evaluate the correctness of the output using a pattern (not an exact ground truth). This paper focuses on functional tests using patterns as evaluation methods.
The fifth metric used in xAI-Bench is remove-and-retrain (ROAR). It involves a re-evaluation of the model, which could itself alter the evaluation.
Another critical factor affecting the scores and the ranking of the algorithms is the data distribution. The authors did circumvent the issue by testing on different distributions. 
However, it remains difficult to decide if the algorithm is failing this specific test or if it is generally sensitive to the data distribution.
xAI-Bench is an excellent initiative to benchmark the correctness of an xAI algorithm, except that it does not allow a clear debugging and does not propose any final ranking of the xAI algorithm to help practitioners and laymen quickly pick the right tool. 
Following the analysis of related work, Section \ref{sec:main} details how our proposed benchmark addresses the highlighted issues.



\section{Compare-xAI}\label{sec:main}
\begin{figure}[h]
 \centering
 \centerline{\includegraphics[width=.9\textwidth]{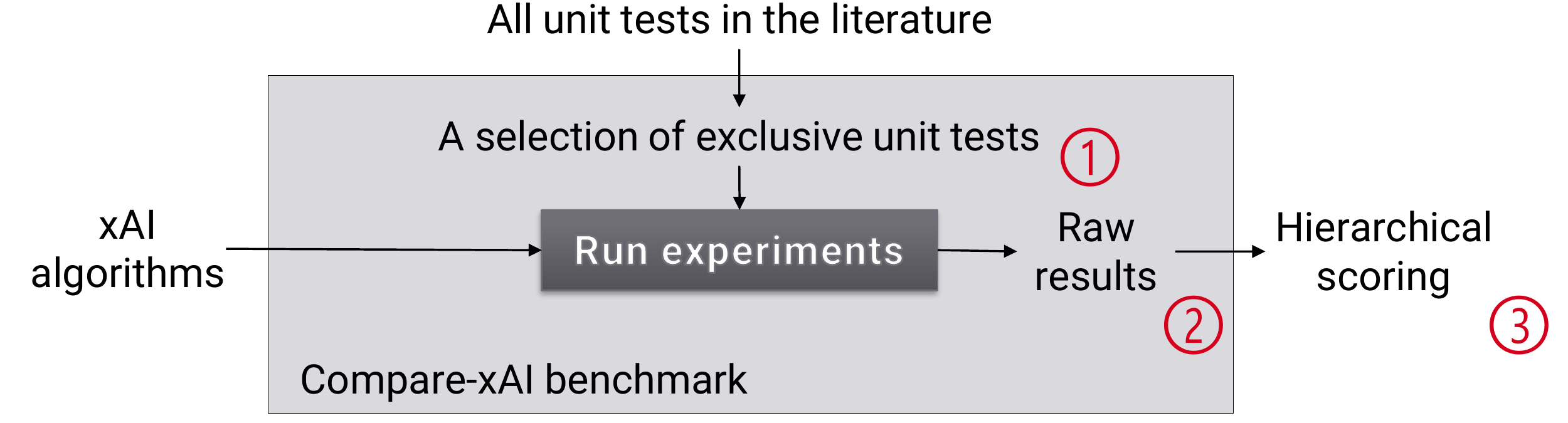}}
 \caption{Compare-xAI's pipeline}
 \label{fig:pipeline}
\end{figure}

Compare-xAI is a quantitative benchmark based solely on functional tests. Compare-xAI is able to evaluate any new xAI algorithm or index any new test. Current proof-of-concept evaluates more than \arabic{nb_explainers} post-hoc xAI algorithms on more than \arabic{nb_tests} functional tests. Compare-xAI outputs a series of scores that allows a multi-dimensional analysis of comparable xAI algorithms. Figure~\ref{fig:pipeline} illustrates Compare-xAI as a pipeline with three added values: First, we collect the known functional tests reported in related literature and filter them according to a clear protocol stated in Section~\ref{sub:test_protocol}. As a second step, each algorithm is tested automatically on each of the collected tests. Experiments follow a protocol detailed in Section~\ref{sub:experiment_protocol}. Each experiment results in one score ranging from 0 (failing) to 1 (succeeding). Compare-xAI has the exclusive advantage of reporting an intermediate score (between 0 and 1) if the algorithm is partially failing the test. Obtained raw results describe an xAI algorithm by \arabic{nb_tests} scores, and it is not easy, at this point, to compare two algorithms. Therefore, raw results are aggregated into a hierarchical scoring method, see Section~\ref{sub:scoring_protocol}.

\subsection{Tests Selection Protocol}
\label{sub:test_protocol}
A functional test consists of a dataset, a model to explain, and an assertion (e.g., an xAI algorithm is expected to detect symmetric input features). 
Section~\ref{sub:tests} highlighted the diversity of tests used in the literature and the importance of filtering inadequate ones. We propose the following rules to ensure a multi-dimensional evaluation of all post-hoc xAI algorithms:
\begin{description}
\item[One end-user requirement per test.] Selected tests should identify and debug a clear end-user requirement within an xAI algorithm. 
This rule ensures that an algorithm will not fail multiple tests for the same reason, and it allows researchers quickly debug the xAI algorithm.
Table~\ref{tab:test_category} regroups a non-exhaustive set of unitary end-user requirements that could alter the algorithm's output and let it fail in explaining the model correctly.
An example of tests sharing the same end-user requirement is ``The failure of the dummy axiom''~\citep{sundararajan2020many} and ``The failure of the linearity axiom''~\citep{sundararajan2020many}: The first test verifies that specific xAI algorithms cannot detect dummy inputs features when the data distribution is modified. The second test verifies linearity between two AI models' output and their explanations. Each test explains how its respective axiom could fail while both share the same root cause: the effect of data distribution on the xAI algorithm. Therefore, Compare-xAI does not implement ``The failure of the linearity axiom''~\citep{sundararajan2020many}. Nevertheless, another test could be found in the literature to verify exactly this axiom.
\item[Undebatable test.] Explanations could target different reasoning. If 2 opposite explanations could be considered correct given the same setup (dataset, AI model), then Compare-xAI would not include such a test. A popular example is the causal explanation of the SHAP values~\citep{janzing2020feature}.


\item[No exact ground truth explanation.] A test contains a scoring function that compares the xAI output to the expected output. The expected output could be an exact set of values, e.g., GT-Shapley's expected output is the Shapley values~\citep{liu2021synthetic}. The expected output could also be a pattern, e.g., given one specific model, feature A's importance should be the highest, or negative, or equal to feature B's. 
Compare-xAI's shortlisted tests rely only on patterns. Protecting this degree of freedom makes this benchmark compatible with different xAI approaches. 
For example, in the case of an adversarial attack test, it is expected that the ranking of the feature importance does not get affected by corrupted models, regardless of the exact ranking proposed. Another good example is the test that assesses the symmetry axiom. Its evaluation function verifies the equality between the feature importance values without having an exact value as a reference. 

\item[Non-redundant tests.] Redundant tests emphasis the failure or the robustness of an xAI. Considering the scoring method of Compare-xAI, redundant tests can not be included. they are manually reported and eliminated from the selection. 


\item[Only tests proposed in the related work.] The first iteration of the Compare-xAI benchmark is limited to the tests reported in former research papers, as many have been presented, discussed, and heavily criticized in the literature. Compare-xAI takes advantage of this extensive research work to build a consistent benchmark. Identifying not examined end-user requirements and proposing new tests 
is rewarding field of future research.

\item[Categorizing functional tests.] In order to cover a large variety of end-user requirements, we propose to categorize shortlisted functional tests into five common groups~\citep{lakkaraju2019faithful}, see Table~\ref{tab:test_category}. 
\end{description}
\begin{table}
\caption{Samples from the shortlisted functional tests.}
\label{tab:test_category}
\centering
\begin{tabular}{p{.08\textwidth}p{.86\textwidth}}
\toprule
Category     & Grouped functional tests \\
\midrule 

Fidelity  & \textbf{Does the algorithm's output reflect the underlying model ?}\\
 & aka faithfulness~\citep{alvarez2018towards}, consistency~\citep{lundberg2018consistent}\\
 &$\bullet$ Counterexample for the symmetry axiom~\citep{lundberg2018consistent}. \\
 &$\bullet$ Test whether features of different importance are represented correctly~\citep{lundberg2018consistent}. \\
 &$\bullet$ Test the detection of feature interaction based on mathematical terms~\citep{tsang2020does}.\\
 &$\bullet$ Effect of feature product on local explanations~\citep{kumar2020problems}.\\
 &$\bullet$ Test if one-hot encoded features are explained correctly~\citep{amoukou2022accurate}.\\
 &$\bullet$ Test if main terms and interaction terms are evaluated correctly~\citep{tilevik2021interpretable}.\\[2mm]

Fragility & \textbf{Is the explanation result susceptible to malicious corruption?}\\
 &$\bullet$ Adversarial attacks can exploit feature perturbation-based xAI algorithms as a vulnerability to lower the importance of specific features~\citep{lakkaraju2020fool}.\\[2mm]

Stability & \textbf{Is the algorithm's output too sensitive to slight changes in the data or model?}\\
 &$\bullet$ Effect of data distribution: Statistical dependence, non-uniform distribution~\citep{janzing2020feature}\\
 &$\bullet$ Effect of feature correlations~\citep{liu2021synthetic,kaur2020interpreting}\\
 &$\bullet$ Effect of noise in the dataset~\citep{sundararajan2020many}\\
 &$\bullet$ Implementation invariance axiom\\[2mm]


Simplicity & \textbf{Can users look at the explanation and easily reason about the model's behavior?}\\
 & aka Explicitness/Intelligibility~\citep{alvarez2018towards}\\
 &$\bullet$ Counterexample of the dummy axiom~\citep{sundararajan2020many}\\
 &$\bullet$ Counterexample of the linearity axiom~\citep{sundararajan2020many}\\[2mm]

Stress & \textbf{Can the algorithm explain models trained on big data?}\\
 &$\bullet$ Test if the xAI algorithm is sensitive to a high number of word tokens (NLP task)~\citep{lundberg2017unified}.\\
 &$\bullet$ Detect dummy pixels in the MNIST dataset~\citep{covert2020understanding}. \\
 &$\bullet$ Effect of data sparsity~\citep{sundararajan2020many}.\\[2mm] 

\textit{Other} &  \textbf{Remaining metrics are not integrated into the hierarchical scoring system albeit reported in the final dataset.}\\
 &$\bullet$ Portability~\citep{molnar2020interpretable} measures the diversity of models' implementation that an xAI algorithm can explain. Portability is tested implicitly by different tests as each one implements a different model/dataset.\\
 &$\bullet$ The relative execution time is an essential factor in choosing an algorithm, and it is mainly influenced by the stress tests.\\
\bottomrule 
\end{tabular}
\end{table}


\subsection{Experiments Protocol}
\label{sub:experiment_protocol}
An experiment takes one test and one xAI algorithm. First, the test environment is initialized by loading the data and training the model. Then the xAI algorithm is asked to explain the model (Globally or for specific data points). Finally, the explanation is compared to the correct answer, and one final score, a real number between 0 and 1, is returned.

It might seem to end-users that the stated patterns in Table~\ref{tab:test_category} are self-evident and verified for every xAI algorithm. As a matter of fact, the public availability and wide usage of particular xAI tools ``swayed several participants to trust the tools without fully understanding them''~\citep{kaur2020interpreting}. 
For this reason, we score each xAI algorithm by following the most common usage of the xAI algorithms. This policy induces the following rules:

\begin{description}
\item[Experiments will only be run once.]
Recent research revealed ``a misalignment between data scientists’ understanding of interpretability tools and these tools’ intended use. Participants misused the tools (either over- or under-used them).''~\citep{kaur2020interpreting}. Compare-xAI is intended for this former group, that is, data scientists not running complementary experiments or repeating the same experiment to test the effect of the noise. Targeting the remaining data scientist, i.e., experts with advanced knowledge of the tools' limits, is left for future work.

\item[No fixed seed for random number generators.] For the same reasons stated above, the seed is not fixed. Shortlisted tests' underlying noise does not hinder any xAI method from correctly explaining a model, i.e., initialized models always learn the tested patterns independently of the seed.

\item[No parameter tuning.] Compare-xAI evaluates algorithms using their default parameters for all tests. Nevertheless, certain xAI algorithm adapts their parameters internally given each task by relying on the model's structure, the dataset size, and the ML task. Around half of the indexed algorithms have at least one binary parameter, and the performance of some would vary with parameter tuning. As there are no precise methods or public tools to fine-tune parameters of xAI algorithms, we suppose the usage of default parameters, by end-users, to be a common misuse of the xAI algorithm. It is important that Compare-xAI reproduces this behavior in order to calculate the comprehensibility score in practice, that is, the ease of correctly interpreting the algorithm's output, considering common practice. 
\end{description}

\subsection{Scoring Protocol}
\label{sub:scoring_protocol}
Compare-xAI's tests result in a set of scores per algorithm, which we call ``raw'' results. A complete comparison between two algorithms A and B should consider the following four points:\newline

\textbf{(1)}A‘s score for ea ch test is greater than or equal to B‘s score;\newline
\textbf{(2)} For each test, A’s execution time is less than or equal to B's execution time;\newline
\textbf{(3)}\label{requirement3}  B's supported models are a subset of A’s (see the portability metric definition in Table~\ref{tab:test_category}); and\newline
\textbf{(4)}\label{requirement4} B's supported output explanations are a subset of A’s.

As a consequence, sorting xAI algorithms by performance is a multi-metric ranking problem. Comparing even two algorithms using this property is impossible, especially with a considerable number of tests. This formulation of the challenge outlines the difficulty faced by practitioners/laymen who are looking for a quick explanation of their models but cannot decide on which xAI algorithm to pick. This challenge is addressed by proposing a relaxation of the scoring dimensions.

\begin{figure}
\centering
\centerline{\includegraphics[width=\textwidth]{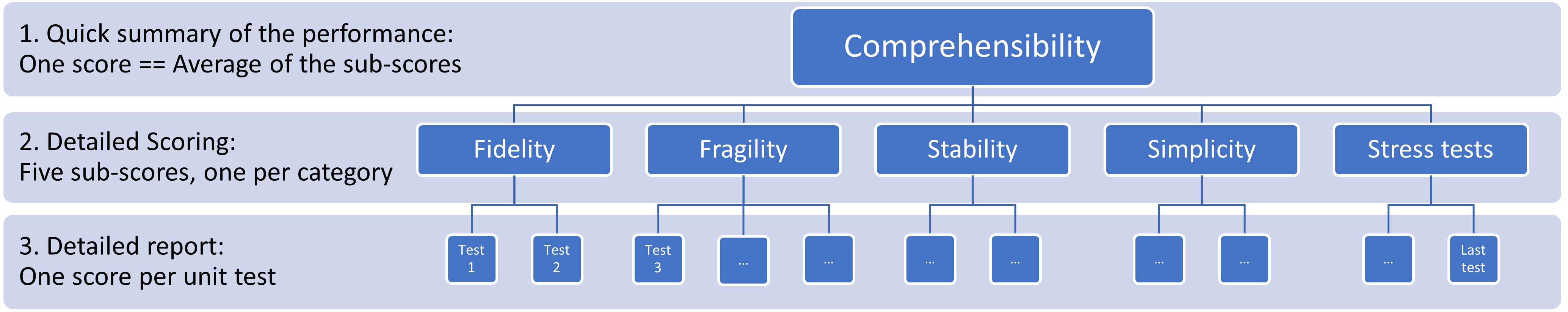}}
\caption{Hierarchical scoring}
\label{fig:scoring}
\end{figure}

First, let us suppose that each end-user is comparing a subset of xAI algorithms which are all able to explain the model of his/her interest, and they all offer the type of explanation required by the end-user (feature importance, attribution, interaction, etc.). On this account, requirements \hyperref[requirement3]{\textbf{(3)}} and \hyperref[requirement4]{\textbf{(4)}}, stated above, might be skipped. 
Second, looking at the overall distribution of the scores, There is no state-of-the-art algorithm that succeeds in all tests. Thus, we opt to promote the ``on-average better explanation'' comparison method. 
\paragraph{Hierarchical scoring.} Figure~\ref{fig:scoring} explains the hierarchical scoring proposed to the end-user to simplify the comparison between xAI algorithms. Level three, the last level in the hierarchical scoring, represents the most detailed report (one score per test) used in the classical ranking method. To simplify it, level two regroups tests per category to propose five scores described in Table~\ref{tab:test_category}. Finally, the first level aggregates the scores into one value, which we named the comprehensibility score.
\paragraph{Comprehensibility score.} Comprehensibility is defined in the literature as ``how much effort is needed for a human to interpret a model correctly?''~\cite{lecue2022tutorial}. Comprehensibility was used as a qualitative metric in related work. To quantify the Comprehensibility metric, we reformulate the definition as follows: 
For an AI model and an xAI algorithm, the effort needed for an end-user to interpret a model correctly is represented by the number of operations he should perform to obtain a correct explanation from the xAI. 
An operation could be, for example, multiple runs of the xAI algorithm to obtain a stable average explanation; a check/edition of the test dataset to adapt the distribution to the xAI's needs; or to perform additional checks against potential adversarial attacks exploiting the xAI algorithm's vulnerabilities.
We quantify the Comprehensibility metric by linking the number of operations to the number of failed functional tests. In other words, if the xAI fails a specific test, the end-user will have to perform additional operations to obtain a correct explanation.
We calculate the Comprehensibility metric as the average over the five subcategories of the second level. Therefore, comparing xAI algorithms becomes more accessible using the Comprehensibility score and the average execution time.

Finally, the evaluation's result is reduced to the comprehensibility score and the average execution time. Remains the dilemma of choosing the fastest algorithm or the most ``comprehensible'' one from the Pareto front. Depending on his/her level of expertise, the end-user is invited to consult any of the three scoring methods made available via the web user interface.

\section{Visualization of the Benchmark}\label{sec:interface}
\begin{figure}
 \centering
\centerline{\includegraphics[width=.85\textwidth]{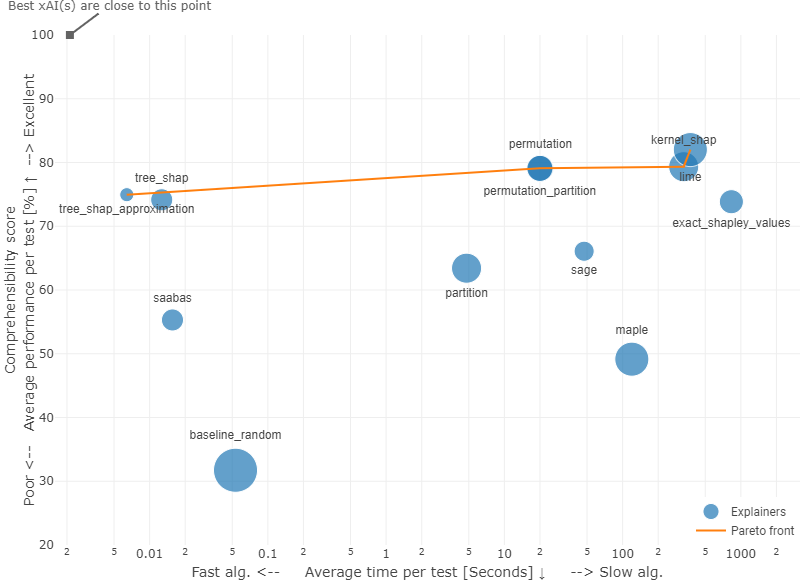}}
 \caption{Global overview of the benchmark}
 \label{fig:benchmark}
\end{figure}

\begin{figure}
 \centering
\centerline{\includegraphics[width=0.95\textwidth]{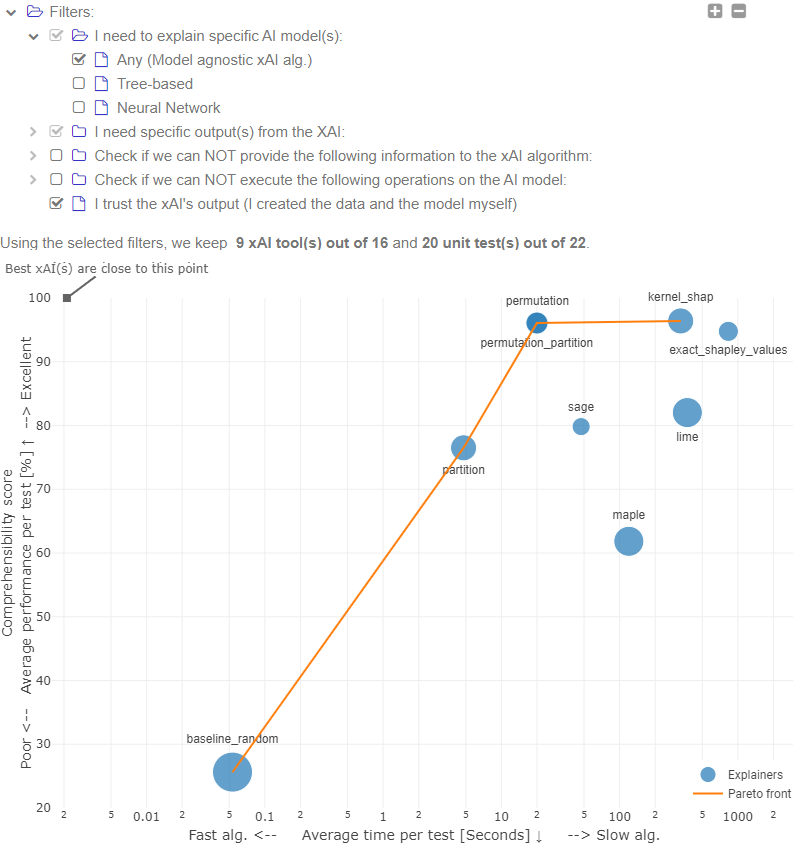}}
 \caption{Benchmark of model-agnostic xAI algorithms outputting feature importance}
 \label{fig:benchmark_restricted}
\end{figure}

For the proof of concept, we consider a small set of popular xAI methods, implement a user interface to easily explore the benchmark results, and make the results publicly available\footnote{ \url{https://karim-53.github.io/cxai/}}. 
Figures~\ref{fig:benchmark} and \ref{fig:benchmark_restricted} do not represent a final benchmark as the set of tests and xAI algorithms are constantly updated. In this section, the demonstration will focus on feature importance. Nevertheless, the benchmark remains adaptable to many forms of post-hoc xAI methods.

Figure~\ref{fig:benchmark} summarizes the performance of the considered set of xAI methods. The best algorithm has the lowest execution time, highest score, and highest portability (bigger dot size). The Pareto front regroups the closest algorithms to the perfect one.
End-users could use the filters on the web interface to describe a specific use case: Figure~\ref{fig:benchmark_restricted} is restricted to the set of model-agnostic xAI algorithms that output (at least) global feature importance. Available filters help the end-user quickly and accurately navigate the massive amount of undocumented properties of available xAI tools.

\begin{figure}
 \centering
\centerline{\includegraphics[width=.85\textwidth]{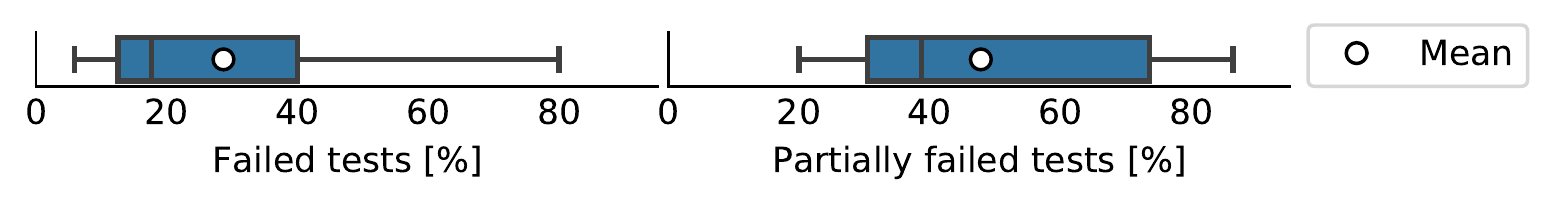}}
 \caption{Box plot of test status per xAI algorithm}
 \label{fig:boxplot}
\end{figure}

At this stage, the end-user picks an xAI algorithm that matches his expected performance and execution time requirements. The detailed report is accessible by clicking on the blue dot representing the algorithm: the end-user gets access to information about the supported AI models, the output of the xAI algorithm, additional information required by the xAI algorithm to run correctly, and essentially the score obtained on each test. The detailed report helps the end-user quickly understand the limit of the xAI algorithm before using it.
Finally, the end-user can compare multiple xAI algorithms given a set of tests that reflect his usage of AI. 
Please keep in mind that selected screenshots are for demonstration purposes only and we are not going to discuss individual scores, but only the benchmark itself.
\paragraph{Discussion}
Compare-xAI is a dynamic benchmark, and its results evolve with the filtered tests/xAI algorithms. Therefore a global analysis of the results is performed without reporting any specific number. 
First, considering all implemented unit tests without filtering, none of the xAI algorithms did obtain the perfect score. Second, obtained comprehensibility scores are very close (50\% of the xAI algorithms obtained a comprehensibility score between 0.70 and 0.85). This clustering reflects the original structure of these xAI algorithms as the majority are permutation-based algorithms. At this level of the analysis, the end-user did save a huge amount of time by understanding which algorithms are almost equivalent and which are relatively faster / explaining better. 
Same for the scores of the second level, analyzing the five scores quickly locate the weak point of a chosen xAI algorithm, since the average difference between the smallest and biggest score is 71.9\% ($\pm31.4 \%$).
Finally, end-users can check the detailed report (level 3). They will mainly go through the failed tests (score < 0.05) or partially failed tests (0.05 $\leq$ score < 0.95). On average, an xAI algorithm is eligible to 12.3 tests ($\pm$ 5) depending on its portability. The distribution of failed tests is skewed, see Figure~\ref{fig:boxplot}. Thus, the median is more representative. The median percentages of failed tests and partially failed tests are 17.6\% and 38.7\%, respectively. Thus, checking the detailed report is expected to be fast.

\section{Limitations and Future Work}\label{sec:limitation}
Compare-xAI's weaknesses are classified into design-related and implementation-related limitations.
\paragraph{Design-related limitations.} 
Compare-xAI is a benchmark made exclusively of quantitative metrics. It is objective as it does not include tests based on human evaluation. A common example from is the study of the human mental model like the investigation of users' preferred explanation style~\citep{jeyakumar2020can}). Another example is the study of the information overload, e.g., xAI's additional output information like the confidence interval~\citep{covert2020understanding}. Mainly, empirical studies are challenging to quantify~\citep{miller2019explanation} and integrate into the comprehensibility score.
These and other non-quantifiable advantages/disadvantages will be included in the description of the xAI algorithm, in the future.
\paragraph{implementation-related limitations.} 
The provided proof of concept includes, for now, \arabic{nb_tests} tests. Currently, none of them cover RL, GAN, or unsupervised learning tasks. The tested form of output is also limited to feature importance (global explanation). Testing feature interaction is still under development. 

Despite the stated limitations, Compare-xAI should fulfill its primary objectives: first, helping laymen pick the right xAI method, and second, helping researchers, practitioners, and laymen avoid common mistakes in interpreting its output.



\section{Conclusion}
\label{sec:conclusion} 


Explaining AI is a delicate task, and some end-users are prone to misuse dedicated tools~\citep{kaur2020interpreting}.
We propose Compare-xAI a unified benchmark indexing +\arabic{nb_explainers} post-hoc xAI algorithms, +\arabic{nb_tests} tests, and +\arabic{nb_papers} research papers.
Compare-xAI reproduces experiments following a selection protocol that highlights the contrast between the theoretical claims of the authors in a paper and the practical implementation offered to the end-user. Selected tests measure diverse properties. The authors did not create any xAI algorithm. Therefore, there is no conflict of interest.
Compare-xAI proposes to deliver the results using an interactive interface as a solution to mitigate human errors in interpreting xAI outputs by making the limits of each method transparent.
Compare-xAI proposes a simple and intuitive scoring method that efficiently absorbs the massive quantity of xAI-related papers.
Finally, Compare-xAI proposes a partial sorting of the xAI methods, toward unifying post-hoc xAI evaluation methods into an interactive and multi-dimensional benchmark.

\section*{Broader Impact}
Compare-xAI is a benchmark with multiple use-cases. It can be seen as a debugging tool for individual xAI algorithms but simultaneously as a global benchmark. Even if Compare-xAI does not offer a total sorting per performance, still it separates comparable algorithms into the Pareto front and the rest.
Compare-xAI allows practitioners to quickly and correctly filter xAI algorithms given their needs and to outline the limitations of the selected ones. 
The end-user, now aware of the limit of the xAI algorithm, would not over-trust the algorithm's output and would avoid common mistakes in explaining a model.
On the other hand, Compare-xAI allows researchers to access a detailed scoring and to answer specific questions such as ``In which case does this xAI algorithm fail?'', ``Is it the only one to solve this issue?'', ``What kind of cases are still not covered by any xAI algorithm?'' etc.
Compare-xAI continuously re-evaluates indexed xAI algorithms to keep an updated benchmark of the state-of-the-art. 
Finally, Compare-xAI is more than a benchmark: it is a comprehensive and standardized related work analysis, while it also works as an evaluation method for new research papers in xAI.

\section*{Acknowledgments}
We thank Dorra El Mekki, Jonas Steinh{\"a}user, Tim Donkiewicz, Marius Daehling, Maximilian Muschalik, Patrick Kolpaczki, and Michael Rapp for their thoughtful feedback on earlier iterations of this work. 

\bibliographystyle{unsrt}
\bibliography{bibliography}


\section*{Checklist}


\begin{enumerate}

\item For all authors...
\begin{enumerate}
 \item Do the main claims made in the abstract and introduction accurately reflect the paper's contributions and scope?
  \answerYes{Compare-xAI is a benchmark for feature importance xAI algorithms. It includes a user interface that helps data scientists quickly run through the scores.}
 \item Did you describe the limitations of your work?
  \answerYes{see~\ref{sec:limitation}}
 \item Did you discuss any potential negative societal impacts of your work?
  \answerNo{We discuss the societal impacts in the ``Broader impact'' section, but we would not qualify it as negative.}
 \item Have you read the ethics review guidelines and ensured that your paper conforms to them?
  \answerYes{}
\end{enumerate}

\item If you are including theoretical results...
\begin{enumerate}
 \item Did you state the full set of assumptions of all theoretical results?
  \answerNA{There are no theoretical results.}
      \item Did you include complete proofs of all theoretical results?
  \answerNA{}
\end{enumerate}

\item If you ran experiments (e.g., for benchmarks)...
\begin{enumerate}
 \item Did you include the code, data, and instructions needed to reproduce the main experimental results (either in the supplemental material or as a URL)?
  \answerYes{ Our code is available at \url{https://github.com/Karim-53/Compare-xAI} and the main readme contains the instructions on how to re-run experiments.}
 \item Did you specify all the training details (e.g., data splits, hyperparameters, how they were chosen)?
  \answerYes{Each test has its own training details defined in its class.}
      \item Did you report error bars (e.g., with respect to the random seed after running experiments multiple times)?
  \answerNo{The protocol of the benchmark reproduces a specific behavior of a layman: results are reported after only one run. We aim to target the second group of end-users (those who run experiments multiple times) in future work by elaborating more on what to report (min, max, or average) and how to keep an equitable comparison between stable and noisy algorithms. Nevertheless, the benchmark stays stable because it averages over multiple tests.}
      \item Did you include the total amount of compute and the type of resources used (e.g., type of GPUs, internal cluster, or cloud provider)?
  \answerYes{See \url{https://github.com/Karim-53/Compare-xAI/blob/main/README.md#23-computing-ressouces=} }
\end{enumerate}

\item If you are using existing assets (e.g., code, data, models) or curating/releasing new assets...
\begin{enumerate}
 \item If your work uses existing assets, did you cite the creators?
  \answerYes{See \url{https://github.com/Karim-53/Compare-xAI/blob/main/README.md#reference=}}
 \item Did you mention the license of the assets?
  \answerNo{}
 \item Did you include any new assets either in the supplemental material or as a URL?
  \answerYes{ \url{https://github.com/Karim-53/Compare-xAI/blob/main/LICENSE} }
 \item Did you discuss whether and how consent was obtained from people whose data you're using/curating?
  \answerNA{We do not survey people.}
 \item Did you discuss whether the data you are using/curating contains personally identifiable information or offensive content?
  \answerNA{}
\end{enumerate}

\item If you used crowdsourcing or conducted research with human subjects...
\begin{enumerate}
 \item Did you include the full text of instructions given to participants and screenshots, if applicable?
  \answerNA{No crowdsourcing.}
 \item Did you describe any potential participant risks, with links to Institutional Review Board (IRB) approvals, if applicable?
  \answerNA{No crowdsourcing.}
 \item Did you include the estimated hourly wage paid to participants and the total amount spent on participant compensation?
  \answerNA{No crowdsourcing.}
\end{enumerate}

\end{enumerate}

\newpage
\appendix
\section{Tests}
For the proof-of-concept, the following list of tests is considered. Note that some tests count twice as they test both feature importance and feature attribution.
\begin{description}

\item[\href{https://github.com/Karim-53/Compare-xAI/tree/main/tests/cough_and_fever.py}{cough\_and\_fever}] answers the following question: \emph{Can the xAI algorithm detect symmetric binary input features?}.
The trained model's equation is [Cough AND Fever]*80.
 The test utilize \textbf{XGBRegressor} model trained on \textbf{a synthetic uniform distribution} dataset (total size: 20000).
 The test procedure is as follows: train a model such that its response to the two features is exactly the same. The xAI algorithm should detect symmetric features (equal values) and allocate them equal importance.
 The score is calculated as follows: 1 if the xAI detect the two features are symmetric. 0 if the difference in importance is above one unit.
 The test is classified in the \textbf{fidelity} category because it is a simple tree model that demonstrate inconsistencies in explanation \citep{lundberg2018consistent}.

\item[\href{https://github.com/Karim-53/Compare-xAI/tree/main/tests/cough_and_fever_10_90.py}{cough\_and\_fever\_10\_90}] answers the following question: \emph{Can the xAI algorithm detect that 'Cough' feature is more important than 'Fever'?}.
The trained model's equation is [Cough AND Fever]*80 + [Cough]*10. Cough should be more important than Fever globally. Locally for the case (Fever = yes, Cough = yes) the feature attribution of Cough should be more important.
 The test utilize \textbf{XGBRegressor} model trained on \textbf{a synthetic uniform distribution} dataset (total size: 20000).
 The test procedure is as follows: train a model with two features with unequal impact on the model. The feature with a higher influence on the output should be detected more important.
 The score is calculated as follows: Return 1 if Cough is more important otherwise 0.
 The test is classified in the \textbf{fidelity} category because it is a simple tree model that demonstrate inconsistencies in explanation due to the tree structure \citep{lundberg2018consistent}.

\item[\href{https://github.com/Karim-53/Compare-xAI/tree/main/tests/x0_plus_x1.py}{x0\_plus\_x1\_distrib\_non\_uniform\_stat\_indep}] answers the following question: \emph{Is the xAI able to explain the model correctly despite a non-uniform distribution of the data?}.
The test demonstrate the effect of data distribution / causal inference.
 The test utilize \textbf{XGBRegressor} model trained on \textbf{a non-uniform and statistically independent} dataset (total size: 10000).
 The test procedure is as follows: Check if the explanation change when the distribution change. Check if non-uniform distributions affect the explanation.
 The score is calculated as follows: returns 1 if the two binary features obtain the same importance.
 The test is classified in the \textbf{stability} category because it assesses the impact of slightly changing the inputs \citep{janzing2020feature}.

\item[\href{https://github.com/Karim-53/Compare-xAI/tree/main/tests/x0_plus_x1.py}{x0\_plus\_x1\_distrib\_uniform\_stat\_dep}] answers the following question: \emph{Is the xAI able to explain the model correctly despite a statistically-dependent distribution of the data?}.
The test demonstrate the effect of data distribution / causal inference. The example was given in both \citep{hooker2021unrestricted} and \citep{janzing2020feature}.
 The test utilize \textbf{XGBRegressor} model trained on \textbf{a uniform and statistically dependent} dataset (total size: 10000).
 The test procedure is as follows: Check if the explanation change when the distribution change. Check if statistically dependent distributions affect the explanation.
 The score is calculated as follows: returns 1 if the two binary features obtain the same importance.
 The test is classified in the \textbf{stability} category because To assess the impact of changing the inputs of f... This way, we are able to talk about a hypothetical scenario where the inputs are changed compared to the true features \citep{janzing2020feature}.

\item[\href{https://github.com/Karim-53/Compare-xAI/tree/main/tests/mnist.py}{mnist}] answers the following question: \emph{Is the xAI able to detect all dummy (constant and useless) pixels?}.
The xAI algorithm should detect that important pixels are only in the center of the image.
 The test utilize \textbf{an MLP} model trained on \textbf{the \href{https://www.openml.org/d/554}{MNIST}} dataset (total size: 70000).
 The test procedure is as follows: simply train and explain the MLP model globally for every pixel.
 The score is calculated as follows: Return the ratio of constant pixels detected as dummy divided by the true number of constant pixels.
 The test is classified in the \textbf{stress} category because of the high number of input features. The test is adapted from \citep{covert2020understanding}.

\item[\href{https://github.com/Karim-53/Compare-xAI/tree/main/tests/fooling_perturbation_alg.py}{fooling\_perturbation\_alg}] answers the following question: \emph{Is the xAI affected by an adversarial attack against perturbation-based algorithms?}.
Model-agnostic xAI algorithms that use feature perturbation methods might be vulnerable to this attack. The adversarial attack exploits a vulnerability to lower the feature importance of a specific feature. Setup: Let's begin by examining the COMPAS data set. This data set consists of defendant information from Broward Couty, Florida. Let's suppose that some adversary wants to mask biased or racist behavior on this data set.
 The test utilize \textbf{a custom function} model trained on \textbf{the \href{https://github.com/propublica/compas-analysis/blob/master/compas-scores-two-years.csv}{COMPAS}} dataset (total size: 4629).
 The test procedure is as follows: The xAI algorithms need to explain the following corrupted model (custom function): if the input is from the dataset then the output is from a biased model. if not then the output is from a fair model.
 The score is calculated as follows: Return 1 if Race is the most important feature despite the adversarial attack. Score decreases while its rank decrease.
 The test is classified in the \textbf{fragility} category because fragility includes all adversarial attacks \citep{ghorbani2019interpretation}.

\item[\href{https://github.com/Karim-53/Compare-xAI/tree/main/tests/dummy_axiom.py}{counterexample\_dummy\_axiom}] answers the following question: \emph{Is the xAI able to detect unused input features?}.
This is a counter example used in literature to verify that SHAP CES do not satisfy the dummy axiom while BSHAP succeed in this test.
 The test utilize \textbf{a custom function} model trained on \textbf{a synthetic} dataset (total size: 20000).
 The test procedure is as follows: Train a model with one extra feature B that is dummy.
 The score is calculated as follows: returns 1 if the dummy feature B obtain a null importance.
 The test is classified in the \textbf{simplicity} category because assigning an importance of zero to dummy feature reflect the model behavior (Fidelity) but also helps the data scientist to quickly understand the model (Simplicity).

\item[\href{https://github.com/Karim-53/Compare-xAI/tree/main/tests/a_and_b_or_c.py}{a\_and\_b\_or\_c}] answers the following question: \emph{Can the xAI algorithm detect that input feature 'A' is more important than 'B' or 'C'?}.
This is a baseline test that the xAI should succeed in all cases. Model: A and (B or C). Goal: make sure that A is more important than B, C. Noise effect: even if the model output is not exactly equal to 1 still we expect the xai to give a correct answer.
 The test utilize \textbf{XGBRegressor} model trained on \textbf{a synthetic} dataset (total size: 20000).
 The test procedure is as follows: The model learns the following equation: A and (B or C). The explanation should prove that A is more important.
 The score is calculated as follows: If A is the most important feature then return 1. If A is the 2nd most important feature then return 0.5 i.e. 1- (1 / nb of feature more important than A).  If A is the last one: return 0 (completely wrong).
 The test is classified in the \textbf{fidelity} category because of the same reason as cough and fever 10-90: A's effect on the output is higher than B or C.

\end{description}

\section{xAI Algorithms}

\begin{description}

\item[\href{https://github.com/Karim-53/Compare-xAI/blob/main/explainers/archipelago.py}{archipelago}] 
 \citep{tsang2020does} 
separate the input features into sets. all features inside a set interact and there is no interaction outside a set. ArchAttribute is an interaction attribution method. ArchDetect is the corresponding interaction detector. 
The xAI algorithm is model agnostic i.e. it can explain any AI model. 
The xAI algorithm can output the following explanations: Feature interaction (local explanation).

\item[\href{https://github.com/Karim-53/Compare-xAI/blob/main/explainers/explainer_superclass.py}{baseline\_random}] 
 \citep{liu2021synthetic} 
Output a random explanation. It is not a real explainer. It helps measure the baseline score and processing time. 
The xAI algorithm is model agnostic i.e. it can explain any AI model. 
The xAI algorithm can output the following explanations: Feature attribution (local explanation), Feature importance (global explanation), Feature interaction (local explanation).

\item[\href{https://github.com/Karim-53/Compare-xAI/blob/main/explainers/shap_explainer.py}{exact\_shapley\_values}] 
 \citep{shapley1953quota} 
is a permutation-based xAI algorithm following a game theory approach: Iteratively Order the features randomly, then add them to the input one at a time following this order, and calculate their expected marginal contribution \citep{sundararajan2020many}. The output is unique given a set of constrains defined in the original paper. 
The xAI algorithm is model agnostic i.e. it can explain any AI model. 
The xAI algorithm can output the following explanations: Feature importance (global explanation). 
The following information are required by the xAI algorithm: 
			 , A reference dataset (input only)
			 , The model's predict function

\item[\href{https://github.com/Karim-53/Compare-xAI/blob/main/explainers/shap_explainer.py}{kernel\_shap}] 
 \citep{lundberg2017unified} 
it approximates the Shapley values with a constant noise \citep{janzing2020feature}. 
The xAI algorithm is model agnostic i.e. it can explain any AI model. 
The xAI algorithm can output the following explanations: Feature attribution (local explanation), Feature importance (global explanation). 
The following information are required by the xAI algorithm: 
			 , A reference dataset (input only)
			 , The model's predict function

\item[\href{https://github.com/Karim-53/Compare-xAI/blob/main/explainers/lime.py}{lime}] 
 \citep{ribeiro2016should} 
it explains the model locally by generating an interpretable model approximating the original one. 
The xAI algorithm is model agnostic i.e. it can explain any AI model. 
The xAI algorithm can output the following explanations: Feature attribution (local explanation), Feature importance (global explanation). 
The following information are required by the xAI algorithm: 
			 , A reference dataset (input only)
			 , The model's predict probability function
			 , Nature of the ML task (regression/classification)
			 , The model's predict function

\item[\href{https://github.com/Karim-53/Compare-xAI/blob/main/explainers/maple.py}{maple}] 
 \citep{plumb2018model} 
is a supervised neighborhood approach that combines ideas from local linear models and ensembles of decision trees \citep{plumb2018model}. 
The xAI algorithm is model agnostic i.e. it can explain any AI model. 
The xAI algorithm can output the following explanations: Feature attribution (local explanation), Feature importance (global explanation). 
The following information are required by the xAI algorithm: 
			 , AI model's structure
			 , A reference dataset (input only)
			 , The train set
			 , The model's predict function

\item[\href{https://github.com/Karim-53/Compare-xAI/blob/main/explainers/shap_explainer.py}{partition}] 
 \citep{lundberg2017unified} 
Partition SHAP approximates the Shapley values using a hierarchy of feature coalitions. 
The xAI algorithm is model agnostic i.e. it can explain any AI model. 
The xAI algorithm can output the following explanations: Feature attribution (local explanation), Feature importance (global explanation). 
The following information are required by the xAI algorithm: 
			 , A reference dataset (input only)
			 , The model's predict function

\item[\href{https://github.com/Karim-53/Compare-xAI/blob/main/explainers/shap_explainer.py}{permutation}] 
 
is a shuffle-based feature importance. It permutes the input data and compares it to the normal prediction 
The xAI algorithm is model agnostic i.e. it can explain any AI model. 
The xAI algorithm can output the following explanations: Feature attribution (local explanation), Feature importance (global explanation). 
The following information are required by the xAI algorithm: 
			 , input features
			 , A reference dataset (input only)
			 , The model's predict function

\item[\href{https://github.com/Karim-53/Compare-xAI/blob/main/explainers/shap_explainer.py}{permutation\_partition}] 
 
is a combination of permutation and partition algorithm from shap. 
The xAI algorithm is model agnostic i.e. it can explain any AI model. 
The xAI algorithm can output the following explanations: Feature attribution (local explanation), Feature importance (global explanation). 
The following information are required by the xAI algorithm: 
			 , input features
			 , A reference dataset (input only)
			 , The model's predict function

\item[\href{https://github.com/Karim-53/Compare-xAI/blob/main/explainers/saabas.py}{saabas}] 
 
explain tree based models by decomposing each prediction into bias and feature contribution components 
The xAI algorithm can explain tree-based models. 
The xAI algorithm can output the following explanations: Feature attribution (local explanation), Feature importance (global explanation). 
The following information are required by the xAI algorithm: 
			 , AI model's structure

\item[\href{https://github.com/Karim-53/Compare-xAI/blob/main/explainers/sage_explainer.py}{sage}] 
 \citep{covert2020understanding} 
Compute feature importance based on Shapley value but faster. The features that are most critical for the model to make good predictions will have large importance and only features that make the model's performance worse will have negative values.

Disadvantage: The convergence of the algorithm depends on 2 parameters: `thres` and `gap`. The algorithm can be trapped in a potential infinite loop if we do not fine tune them. 
The xAI algorithm is model agnostic i.e. it can explain any AI model. 
The xAI algorithm can output the following explanations: Feature importance (global explanation). 
The following information are required by the xAI algorithm: 
			 , True output of the data points to explain
			 , A reference dataset (input only)
			 , The model's predict function

\item[\href{https://github.com/Karim-53/Compare-xAI/blob/main/explainers/shap_interaction.py}{shap\_interaction}] 
 \citep{owen1972multilinear} 
SI: Shapley Interaction Index. 
The xAI algorithm is model agnostic i.e. it can explain any AI model. 
The xAI algorithm can output the following explanations: Feature interaction (local explanation).

\item[\href{https://github.com/Karim-53/Compare-xAI/blob/main/explainers/shapley_taylor_interaction.py}{shapley\_taylor\_interaction}] 
 \citep{sundararajan2020shapley} 
STI: Shapley Taylor Interaction Index. 
The xAI algorithm is model agnostic i.e. it can explain any AI model. 
The xAI algorithm can output the following explanations: Feature interaction (local explanation).

\item[\href{https://github.com/Karim-53/Compare-xAI/blob/main/explainers/shap_explainer.py}{tree\_shap}] 
 \citep{lundberg2018consistent}
accurately compute the shap values using the structure of the tree model. 
The xAI algorithm can explain tree-based models. 
The xAI algorithm can output the following explanations: Feature attribution (local explanation), Feature importance (global explanation). 
The following information are required by the xAI algorithm: 
			 , AI model's structure
			 , A reference dataset (input only)

\item[\href{https://github.com/Karim-53/Compare-xAI/blob/main/explainers/shap_explainer.py}{tree\_shap\_approximation}] 
 
is a faster implementation of shap reserved for tree based models. 
The xAI algorithm can explain tree-based models. 
The xAI algorithm can output the following explanations: Feature attribution (local explanation), Feature importance (global explanation). 
The following information are required by the xAI algorithm: 
			 , AI model's structure
			 , A reference dataset (input only)

\end{description}

\section{Test Results}

Table \ref{tab:scores} contains test results without using any filter. 
\textbf{Tests} is the number of completed tests.
\textbf{Time} is the average execution time per test. It informs the user about the relative difference in execution time between algorithms.

\begin{table}[h]
\caption{Sample of the benchmark scores.}
\label{tab:scores}
\centering
\begin{tabular}{lrr|rrrrr|r} 
\toprule
xAI algorithm & Time & Tests & Fidelity & Fragility & Stability & Simplicity & Stress test & Comprehen-\\
              & [Seconds]  &  & [\%] & [\%] & [\%] & [\%] & [\%] &sibility [\%]\\
\midrule
baseline\_random         & $<1$ & 22 & 12.9 & 50.0 & 30.7 & 0 & 33.3 & 31.7\\
exact\_shapley\_values    & 831 & 12 & 100.0& 11.1 & 84.3 & 100.0& & 73.9 \\
kernel\_shap             & 328 & 15 & 100.0& 11.1 & 85.6 & 100.0& 100.0& 79.3\\
lime                    & 373 & 17 & 89.0 & 0 & 99.7 & 98.5 & 40.9 &82.0 \\
maple                   & 119 & 17 & 60.0 & 11.1 & 100.0& 0 & 25.5 & 49.2\\
partition               & 5 & 15 & 100.0& 11.1 & 84.3 & 100.0& 21.7 & 63.4 \\
permutation             & 20 & 13 & 100.0& 11.1 & 84.3 & 100.0& 100.0& 79.1 \\
permutation\_partition   & 20 & 13 & 100.0& 11.1 & 84.3 & 100.0& 100.0& 79.1  \\
saabas                  & $<1$ & 11 & 60.0 &  & 50.6 &  &  & 55.1 \\
sage                    & 47 & 10 & 66.7 & 11.1 & 96.9 & 100.0& 55.7 & 66.1 \\
tree\_shap               & $<1$ & 11 & 64.0 &  & 84.3 &  &  &74.2\\
tree\_shap\_approximation & $<1$ & 7 & 100.0&  & 49.9 &  &  &74.2\\
\bottomrule 
\end{tabular}
\end{table}

\end{document}